\documentclass[aps,prl,twocolumn,superscriptaddress]{revtex4}
\usepackage{graphicx,amssymb}
\usepackage{bm}

\newcommand\CuFeGeO{Cu$_2$Fe$_2$Ge$_4$O$_{13}$}

\newcommand\Cuion{Cu$^{2+}$}

\begin{document}


\title{Cooperative ordering of gapped and gapless spin networks in \CuFeGeO}


\author{T. Masuda}
\email[]{masudat@ornl.gov}
\affiliation{Condensed Matter Science Division, Oak Ridge National
Laboratory, Oak Ridge, TN 37831-6393, USA}

\author{A. Zheludev}
\affiliation{Condensed Matter Science Division, Oak Ridge National
Laboratory, Oak Ridge, TN 37831-6393, USA}

\author{B. Grenier}
\affiliation{CEA-Grenoble, DRFMC / SPSMS / MDN, 17 rue des
Martyrs, 38054 Grenoble Cedex, France}

\author{S. Imai}
\altaffiliation{ Present address: Semiconductor R \& D Center,
Yokohama R \& D Laboratories,
 The Furukawa Electric Co., Ltd. }
\affiliation{Department of Advanced Materials Science, University
of Tokyo, 5-1-5, Kashiwa-no-ha, Kashiwa, 277-8581, Japan}

\author{K. Uchinokura}
\altaffiliation{ Present address: The Institute of Physical and
Chemical Research (RIKEN), Wako, Saitama 351-0198, Japan}
\affiliation{Department of Advanced Materials Science, University
of Tokyo, 5-1-5, Kashiwa-no-ha, Kashiwa, 277-8581, Japan}

\author{E. Ressouche}
\affiliation{CEA-Grenoble, DRFMC / SPSMS / MDN, 17 rue des
Martyrs, 38054 Grenoble Cedex, France}

\author{S. Park}
\affiliation{NIST Center for Neutron Research, National Institute
of Standards and Technology, Gaithersburg, Maryland 20899, USA}


\date{\today}

\begin{abstract}
The unusual magnetic properties of a novel low-dimensional quantum
ferrimagnet \CuFeGeO\ are studied using bulk methods, neutron
diffraction and inelastic neutron scattering. It is shown  that
this material can be described in terms of two low-dimensional
quantum spin subsystems, one gapped and the other gapless,
characterized by two distinct energy scales. Long-range magnetic
ordering observed at low temperatures is a cooperative phenomenon
caused by weak coupling of these two spin networks.
\end{abstract}

\pacs{75.10.Jm, 75.25.+z, 75.50.Ee}

\maketitle

In low-dimensional magnets quantum spin fluctuations are often
potent enough to completely destroy long range order even at zero
temperature. Quantum disorder in systems with a spin gap, also
known as ``quantum spin liquids'', is particularly robust. These
systems resist ``spin freezing'', i.e., forming a magnetically
ordered state, even in the presence of weak 3-dimensional (3D)
interactions or magnetic anisotropy. In contrast, gapless low
dimensional magnets have a quantum-critical ground state and,
being extremely sensitive to any external perturbations, can be
forced to order more easily. Either way, when
quasi-low-dimensional magnets {\it do} order, it typically
involves some interesting and exotic physics. For example, spin
freezing may be induced by {\it non-magnetic}
\cite{Hase93b,Uchiyama99} or magnetic \cite{Zheludev01a}
impurities, or by a Bose-condensation of magnons induced by an
external magnetic field \cite{Honda98a,Oosawa99a,Chen01a}.

An alternative ordering mechanism can be envisioned in a complex
material that incorporates two intercalated low-dimensional
subsystems. Particularly interesting are cases involving a mixture
of gapless and gapped entities. If these two spin networks were
totally isolated from one another, they would remain disordered at
any $T>0$, due to their reduced dimensionality~\cite{Mermin66}.
However, arbitrary weak interactions between the two subsystems
can drastically change this outcome. Divergent one- or
two-dimensional correlations in the gapless subsystem, combined
with short-range correlations in the third direction propagated by
the subsystem with a gap, will ultimately coalesce into a 3D
ordered phase at a finite temperature. Provided that interactions
between the two subsystems are relatively weak, this new {\it
cooperative ordered state} can be expected to possess some rather
unique characteristics. Among these are a strong quantum
suppression of ordered moments in the nominally gapped subsystem,
a coexistence of conventional spin waves and remanent gap modes,
interactions and ``anticrossing'' between such excitations, {\it
etc}. In the present work we report an experimental realization
and study of this novel cooperative ordering mechanism in the
quantum ferrimagnet \CuFeGeO.

\begin{figure}
\begin{center}
\includegraphics[width=8.5cm]{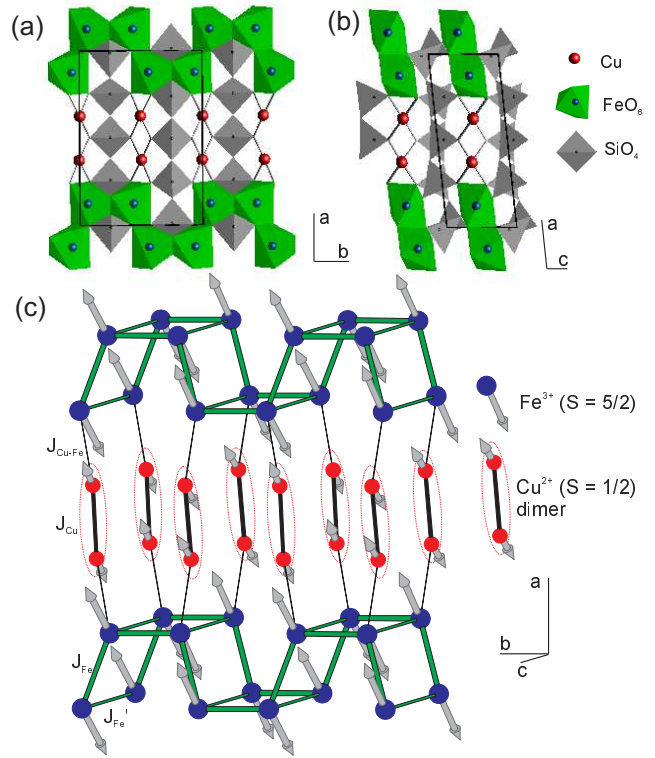}
\end{center}
\caption{(a) and (b) Crystal structure of
Cu$_2$Fe$_2$Ge$_4$O$_{13}$ projected onto $(a,b)$ and $(a,c)$
crystallographic planes. (c) Magnetic structure and possible
exchange pathways in Cu$_2$Fe$_2$Ge$_4$O$_{13}$. } \label{fig1}
\end{figure}

The crystal structure of \CuFeGeO\ \cite{Masuda03a} can be viewed
as a modification of that of the famous spin-Peierls compound
CuGeO$_3$~\cite{Hase93a}. It consists of three distinct
components: ``crankshaft''-shaped chains of edge sharing FeO$_6$
octahedra, Cu-O-Cu dimers, and oligomers of four GeO$_4$
tetrahedra~\cite{Masuda03a}. The mutual arrangement of these
structural elements is visualized in Fig.~\ref{fig1} (a) and (b).
The space group is monoclinic, $P2_1/m$, with lattice parameters
$a$ = 12.101, $b$ = 8.497, $c$ = 4.869, and $\beta =
96.131^{\circ}$. Fe$^{3+}$ and Cu$^{2+}$ carry $S$ = 5/2 and $S$ =
1/2, respectively. The crankshaft FeO$_6$ chains run along the $b$
axis. They are linked by GeO$_4$ tetrahedra in $c$ direction and
by \Cuion-dimers in the $a$ direction. A plausible network of
superexchange pathways is shown in Fig.~\ref{fig1} (c). The key
piece of information required to understand the magnetism of
\CuFeGeO\ is the hierarchy of the corresponding exchange
constants. Knowing this hierarchy will allow us to formulate the
model in terms of spin ``chains'', ``planes'' and ``dimers''.



\begin{figure}
\begin{center}
\includegraphics[width=8.7cm]{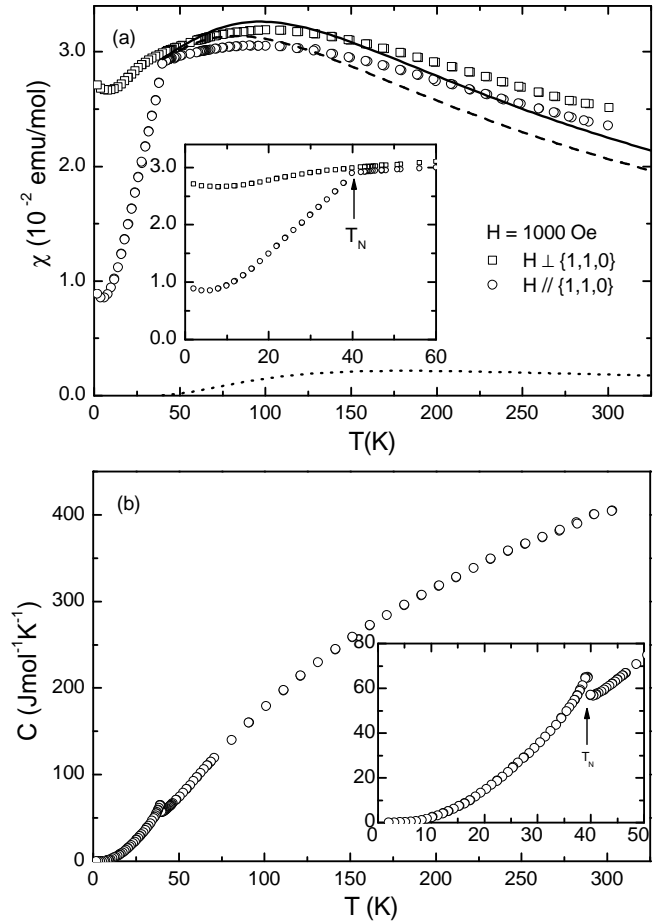}
\end{center}
\caption{(a) Magnetic susceptibility of \protect\CuFeGeO\ measured
in field applied perpendicular and parallel to cleaved plane
(symbols). Solid, dashed and dotted lines are explained in the
text. (b) Heat capacity of \protect\CuFeGeO\ in zero field. }
\label{fig2}
\end{figure}

Single crystals of \CuFeGeO ~were grown in O$_2$ atmosphere by
floating zone method and were characterized using bulk magnetic
susceptibility and heat capacity measurements. The $\chi(T)$
curves measured using a commercial SQUID magnetometer in an
$H=1000$~Oe magnetic field applied perpendicular and parallel to
\{1 1 0\} plane are plotted in open symbols in Fig.~\ref{fig2}
(a). These data are characteristic of a quasi-low-dimensional
antiferronmagnet: the development of broad maximum at $T \sim 100$
K is followed by an antiferromagnetic phase transition at
$T_\mathrm{N}=39$~K. This transition is also manifest in the heat
capacity data shown in Fig.~\ref{fig2} (b), and is the only
prominent feature in our temperature range.

As a ``zeroth approximation'', the measured susceptibility was fit
to a sum of two contributions calculated for isolated $S=5/2$
chains~\cite{Fisher64} and $S=1/2$ dimers. The best fit is
obtained assuming the Cu-Cu and Fe-Fe exchange constants to be
$J_{\rm Cu} = 25$~meV and $J_{\rm Fe}=1.7$~meV, respectively. The
average gyromagnetic ratio of Cu$^{2+}$ and Fe$^{3+}$ is
determined to be $g=2.04$ by room-temperature ESR measurements.
The result of the fit is shown in a solid line in Fig.~\ref{fig2},
with the dashed and dotted lines representing the Fe-chain and
Cu-dimer components, correspondingly. Though rather crude, this
simple analysis gives us a rough idea of the magnitude of the
relevant exchange constants.

\begin{figure}
\begin{center}
\includegraphics[width=8.7cm]{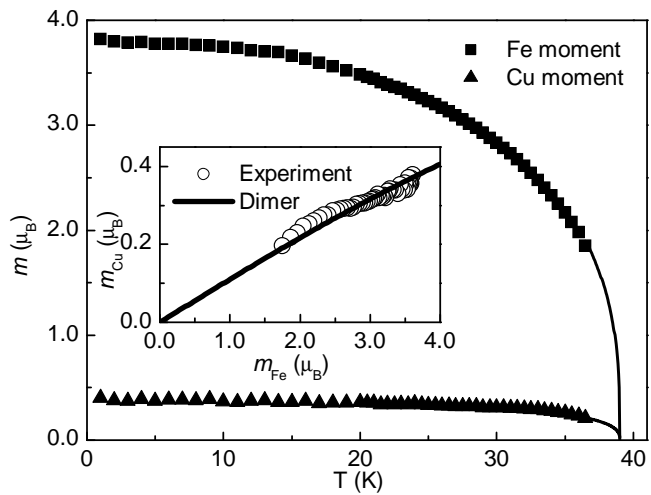}
\end{center}
\caption{Measured temperature dependence of the Fe$^{3+}$- and
Cu$^{2+}$- ordered moments (main panel). Inset: ordered moment on
the Cu$^{2+}$-sites plotted vs. that on Fe$^{3+}$. The solid line
is a fit to the staggered magnetization curve for an isolated
dimer} \label{fig3}
\end{figure}

The spin arrangement in the magnetically ordered state was
determined by neutron diffraction experiments on the CRG-D23
lifting-counter diffractometer at Institute of Laue Langevin. 338
nuclear and 222 magnetic ( propagation vector $(H, K, 0.5)$ )
independent reflections were collected at $T=1.5$~K using a 0.3~g
sample. The results unambiguously indicate an almost antiparallel
and slightly canted spin structure with all spins predominantly in
$(a,c)$ crystallographic plane, as shown in Fig.~\ref{fig1} (c).
The most important result pertains to the magnitudes of the
ordered moments: $m_{\rm Cu} = 0.38(4) \mu _B$ and $m_{\rm Fe} =
3.62(3) \mu _B$. The moment's orientations in the standard
Cartesian coordinate system linked to the crystallographic unit
cell were found to be ${\bm m}_{\rm Cu} = (-0.227,~ 0.035,~
-0.301) \mu _B$ and ${\bm m}_{\rm Fe} = (2.163,~ 0.121,~ 2.892)
\mu _B$, respectively. The temperature dependence of the magnetic
structure was deduced from the behavior of $(2~ 1~ -0.5)$, $(0~ 2~
0.5)$, $(-1~ 1~ 0.5)$ and $(-2~ 1~ 1.5)$ Bragg reflections. The
evolution of sublattice moments for Fe$^{3+}$ and Cu$^{2+}$ are
shown in the main pannel of Fig.~\ref{fig3}. They were determined
under the assumption that all spin orientations are
$T$-independent.

The observed antiparallel spin arrangement indicates that all the
exchange constants in our model for \CuFeGeO\ are
antiferromagnetic (Fig.~\ref{fig1}). Our main finding, however, is
the drastic suppression of the ordered moments on the Cu$^{2+}$
sites, as compared to their classical expectation value of
$1\mu_{\mathrm{B}}$. To explain this behavior we have to assume
that the Cu-dimers are effectively isolated from the rest of the
system, and become only partially polarized by the long-range
order on the Fe-sites. In other words, $J_{\rm Cu} \gg J_{\rm
Cu-Fe}$, and the Cu$^{2+}$ moments are reduced due to strong {\it
local} quantum spin fluctuations within the dimers. The
suppression of the Fe$^{3+}$ moment is less pronounced. This is
undoubtedly due to the large value of the Fe$^{3+}$ spin, and
allows us to regard the Fe-subsystem as a semiclassical one. The
observed small canting of the spin structure can be attributed to
Dzyaloshinskii-Moriya interactions and other insignificant
anisotropy effects.

The strongly dimerized nature of the Cu$^{2+}$ spins has one
interesting consequence. At the Mean Field (MF) level, if each
dimer is treated as single inseparable entitiy, the ordered moment
on the Cu-sites is seen as being {\it induced} by an effective
external staggered exchange field. It is determined by the
staggered magnetization function $M_{+-}(H_{+-},T)$ for an
isolated dimer. Excluding direct inter-dimer interactions, the
exchange field $H_{+-}$ is proportional to the ordered moment on
the Fe-subsystem, {\it regardless} of the internal network of the
latter. Assuming that $k_\mathrm{B}T\ll J_{\rm Cu}$,
$M_{+-}(H_{+-},T)\approx M_{+-}(H_{+-},0)\equiv
M^{(0)}_{+-}(H_{+-})$. This establishes a unique relation between
$m_\mathrm{Cu}$ and $m_\mathrm{Fe}$:
\begin{equation}
m_\mathrm{Cu}= M^{(0)}_{+-}(\alpha m_\mathrm{Fe}),\label{lin}
\end{equation}
where $\alpha\sim J_\mathrm{Cu-Fe}$ is the effective MF coupling
constant. For \CuFeGeO\ this relation indeed holds very well, as
shown in the inset of Fig.~\ref{fig3}. Here the solid line is a
best fit of Eq.~(\ref{lin}) to the experimental data obtained with
$J_\mathrm{Cu-Fe}/J_\mathrm{Cu}=0.11$.

While the Cu-subsystem appears to be well described as a
collection of dimers, the network of Fe$^{3+}$ spins is
considerably more complicated, with several possible interaction
pathways in the $(b,c)$ plane. To better understand these
interactions we performed a series of inelastic neutron scattering
experiments that probed the low-energy spin excitation spectrum.
The data were collected for energy transfers of up to 10~meV using
the SPINS cold 3-axis spectrometer installed at the National
Institute of Standards and Technology and the HB1 thermal 3-axis
spectrometer at the High Flux Isotope Reactor at Oak Ridge
National Laboratory. Two co-aligned single crystal samples had a
total mass of 3.5~g. Several configurations were exploited, with
final neutron energies of 5~meV, 3~meV or 13.5~meV, with several
settings of the sample relative to the scattering plane. A
detailed report of these measurements will be published elsewhere,
and only the main results will be summarized here.

\begin{figure}
\begin{center}
\includegraphics[width=8.7cm]{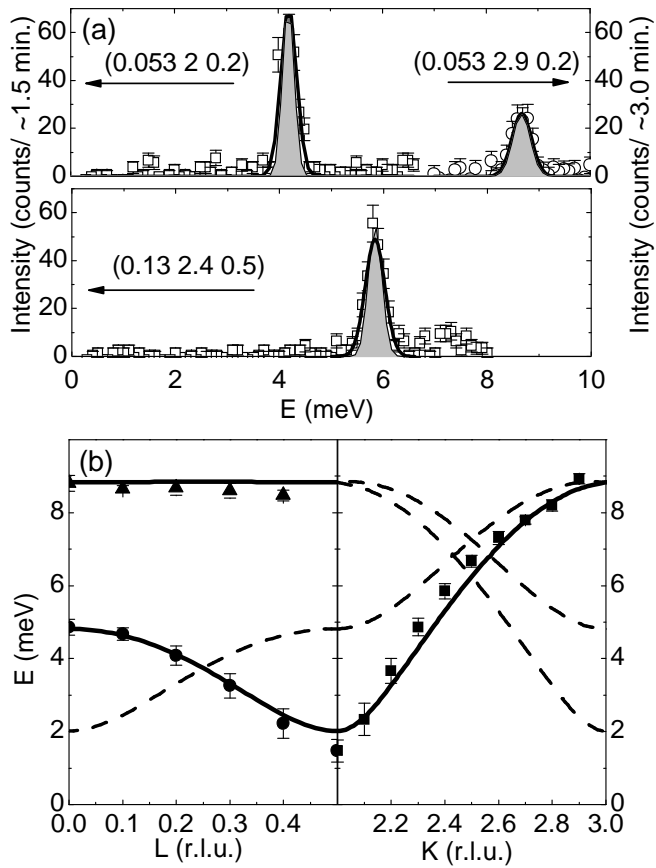}
\end{center}
\caption{(a) Typical constant-$q$ scans collected in \CuFeGeO\ at
$T = 1.4$~K for different wave vectors. Shaded Gaussian curves
represent the experimental energy resolution. Solid lines are
Gaussian fits to the data. (b) Measured dispersion of low-energy
spin waves in \CuFeGeO\ (symbols). Lines represent  spin wave
theory fits to the data as described in the text. } \label{fig4}
\end{figure}

Sharp spin wave excitations of resolution-limited energy width
were observed at all wave vectors, as illustrated by typical scans
shown in Fig.~\ref{fig4} (b). At least two separate excitation
branches are present. Their dispersion along the $a^{\ast}$
direction was undetectable in our experiments. In contrast, the
dispersion along the $b^{\ast}$ and $c^{\ast}$ reciprocal-space
axes is quite pronounced.  The measured dispersion curves are
plotted in symbols in Fig.~\ref{fig4} (a). These low-energy data
can be explained by conventional spin wave theory for a
2-dimensional network that incudes only the Fe$^{3+}$ spins, and
direct Fe-Fe exchange interactions shown in
Fig.~\ref{fig1}~\footnote{Due to a small alternation in bond
length, the coupling constant $J_{\rm Fe}$ may actually slightly
alternate from one bond to the next. This alternation is
apparently too weak to affect the dispersion relations measured in
this work, and was thus not included in the spin wave
calculation}. Dispersion curves for this model were calculated as
explained in Ref.~\cite{Saenz62}. With eight Fe$^{3+}$ ions per
magnetic unit cell there are actually four two-fold degenerate
spin wave branches. Due to particular values of the 3-dimensional
magnetic unit cell structure factors, only two modes produce
strong contributions at a time for each of the scans shown. An
excellent fit to the data is obtained assuming $J_{\rm Fe} =
1.60(2)$~meV and $J_{\rm Fe}' = 0.12(1)$~meV, and assuming a small
empirical anisotropy gap $\Delta=2.02(40)$~meV. The result of the
fit is shown in solid lines in Fig.~\ref{fig4} (a). The obtained
value of $J_{\rm Fe}$ is consistent with our previous estimate
based on simple-minded fits to magnetic susceptibility data. We
conclude that the Fe-layers are indeed to be viewed as chains
running along the $b$ crystallographic axis, with only weak
inter-chain coupled along the $c$ direction.

The fact that the low-energy excitations are entirely due to the
Fe$^{3+}$ moments confirms the main idea behind our
``two-subsystem'' model for \CuFeGeO. The key point is that
$J_\mathrm{Fe-Cu}$, $J_\mathrm{Fe} \ll J_\mathrm{Cu}$. As
previously explained in Ref.~\cite{Zheludev01a} for the case of
R$_2$BaNiO$_5$ nickelates, such an hierarchy of exchange constants
leads to a {\it separation of energy scales} of magnetic
excitations. The Cu-centered (dimer) excitations are confined to
high frequencies, while low frequencies  are entirely dominated by
``acoustic'' fluctuations of the Fe$^{3+}$ spins. However, the
coupling between the two subsystems {\it can not be ignored}: it
is essential for completing a 3D network of magnetic interactions
and enabling long-range magnetic ordering at a finite temperature.
At low frequencies the dimers can be integrated out, and replaced
by an effective interaction $J_\mathrm{eff}$ between Fe$^{3+}$
spins along the $a$ axis. Since the staggered susceptibility of
each Cu-dimer is proportional to $1/J_{\rm Cu}$, we can estimate
$J_\mathrm{eff}\propto J_{\rm Fe-Cu}^2/J_{\rm Cu}$. The absence of
spin wave dispersion along the $a$ axis can thus be explained
either by the weakness of $J_{\rm Fe-Cu}$ that couples the two
subsystems, or by the large value of $J_{Cu}$.

To summarize, our experiments reveal the unconventional {\it
cooperative} ordering of the low-dimensional quantum spin
subsystems of \CuFeGeO, and the persistence of quantum spin
fluctuations even in the 3D-ordered phase. A very interesting
remaining issue is that of Cu$^{2+}$-centered excitations in this
material, expected to occur at higher energy transfers: what
happen to the triplet dimer modes when long-range order sets in?
In the context of separation of energy scales, they may be
expected to survive in the ordered state. In this case, at least
one member of the triplet will have {\it longitudinal}
polarization, and be therefore totally incompatible with
convectional spin wave theory. Another exciting avenue for future
studies are other members of the hypothetical family of compounds
with the general formula
Cu$_{n-2}$Fe$_2$Ge$_n$O$_{3n+1}$~\cite{Masuda03a}. In these
species the Fe layers alternate with layers of Cu-oligomers of
length $n - 2$. Cooperative ordering can be expected in all these
materials, but should be qualitatively different in even-$n$ and
odd-$n$ members, where the Cu-subsystem is either gapped and
gapless, respectively.


We thank Dr. Chakoumakos for crystal structure analysis in the
early stage of this study. Work at ORNL was carried out under
Contracts No. DE-AC05-00OR22725, US Department of Energy.
Experiments at NIST were supported by the NSF through DMR-0086210
and DMR-9986442. Work in Japan was supported in part by a
Grant-in-Aid for COE Research ``SCP coupled system" from the
Ministry of Education, Science, Sports and Culture of Japan.



\end{document}